\documentclass[conference]{IEEEtran}
\IEEEoverridecommandlockouts

\ifCLASSINFOpdf
  \usepackage[pdftex]{graphicx}
  \graphicspath{{./figures/}}
  \DeclareGraphicsExtensions{.pdf}
\else
  \usepackage[dvips]{graphicx}
  \graphicspath{{./figures/}}
  \DeclareGraphicsExtensions{.ps}
\fi

\usepackage{hyperref}
\usepackage{amsmath,amssymb} 
\usepackage{listings}
\usepackage[numbers,sort&compress]{natbib}

\newtheorem{example}{Example}[section]

\begin{document}
\title{SABRE: A Tool for Stochastic Analysis of Biochemical Reaction Networks\thanks{This
research has been partially funded by the Swiss National Science Foundation
under grant 205321-111840 and by the Cluster of Excellence
on Multimodal Computing and Interaction at Saarland University.}}

\author{
\IEEEauthorblockN{Frederic Didier}
\IEEEauthorblockA{School of Computer and \\
Communication Sciences, \\
EPFL, Switzerland
}
\and
\IEEEauthorblockN{Thomas A. Henzinger}
\IEEEauthorblockA{IST Austria \\
Institute of Science and \\
Technology Austria
}
\and
\IEEEauthorblockN{Maria Mateescu}
\IEEEauthorblockA{School of Computer and \\
Communication Sciences, \\
EPFL, Switzerland
}
\and
\IEEEauthorblockN{Verena Wolf}
\IEEEauthorblockA{Department of Computer Science, \\
Saarland University, \\Saarbr\"ucken, Germany
}}

\maketitle

\begin{abstract}
The importance of stochasticity within biological systems has been shown repeatedly
during the last years and has raised the need for efficient stochastic tools.
We present SABRE, a tool for stochastic analysis of biochemical reaction networks.
SABRE implements fast adaptive uniformization (FAU), a direct numerical approximation
algorithm for computing transient solutions of biochemical reaction networks.
Biochemical reactions networks represent biological systems studied at a molecular level
and these reactions can be modeled as transitions of a Markov chain.
SABRE accepts as input the formalism of guarded commands, which it interprets
either as continuous-time or as discrete-time Markov chains.
Besides operating in a stochastic mode, SABRE may also perform a deterministic analysis
by directly computing a mean-field approximation of the system under study.
We illustrate the different functionalities of SABRE by means of biological case studies. 
\end{abstract}

\section{Introduction}

Markov chains are an omnipresent modeling approach in the applied
sciences.  Often, they describe \emph{population processes}, that is,
they operate on a multidimensional discrete state space, where each
dimension of a state represents the number of individuals of a certain
type.  Depending on the application area, ``individuals'' may be
customers in a queuing network, molecules in a chemically reacting
volume, servers in a computer network, actual individuals in a population, etc.

Here, we are particularly interested in dynamical models of
biochemical reaction networks, such as signaling pathways, gene
expression networks, and metabolic networks.  Biochemical reaction
networks operate
on an abstraction level where a state of the system is given by an
$n$-dimensional vector of chemical populations, that is, the system
involves $n$ different types of molecules and the $i$-th element of the state vector
represents the number of molecules of type~$i$.  Molecules collide
randomly and may undergo chemical reactions, which change the state of
the system.  Classical modeling approaches in biochemistry are based
on a system of ordinary differential equations that assume a
continuous \emph{deterministic} change of chemical concentrations.  
During the last decade, \emph{stochastic} analysis of biochemical reaction networks
has seen growing interest because it captures molecular noise~\cite{fedoroff-fontana}, 
which arises from the randomness of the discrete events in the cell.
Molecular noise is of interest because it significantly influences fundamental biological
processes such as gene expression
\cite{elowitz-etal02,Thattai}, decisions of the cell fate\cite{arkin2,Maamar},
and circadian oscillations\cite{Didier,Barkai}.

Within the setting of stochastic analysis, biochemical reaction networks are 
modeled as discrete-state continuous-time Markov processes(CTMC) as suggested by Gillespie
within the theory of stochastic chemical kinetics~\cite{gillespie77}.
The evolution of a CTMC is given by a system of linear ordinary differential 
equations, known as the \emph{chemical master equation} (CME).
A single equation in the CME describes the time-derivative of the probability  
of a certain state at all times $t\ge 0$. Thus,  
the solution of the CME is the probability distribution over all  
states of the CTMC at a particular time $t$, 
that is, the transient state probabilities at time $t$.
The solution of the CME is then used  to derive measures of interest
such as the distribution of
switching delays~\cite{mcadams-arkin97}, the distribution of the time of DNA
replication initiation at different origins~\cite{Patel2006}, or the
distribution of gene expression products~\cite{distr}.  Moreover, many
parameter estimation methods require the computation of the posterior
distribution because means and variances do not provide enough
information to calibrate parameters~\cite{Wilkinsonpaper}.

Statistical estimation procedures such as Monte Carlo simulation are widely used
to estimate the probability distribution of the underlying Markov process, 
because for realistic systems the size of the CME is very large or even infinite, and 
numerical methods become infeasible. 
Several tools for Monte Carlo simulation  have been developed~\cite{dizzy, Heiner08snoopy, copasi}.
Recent work, however, indicates that numerical approximation
methods for the CME can be used to compute the transient state probabilities
more accurately and, depending on the measures of interest, 
with shorter running times~\cite{CMSB09}.
Especially if the probabilities of interest are small, numerical approximations
turn out to be superior to Monte Carlo simulation, because the later requires 
a large number of simulation runs in order to bound the statistical error appropriately.  
For estimating event probabilities, a higher precision level is necessary
than for estimating cumulative measures such as expectations, and
simulation based methods have a slow convergence because doubling the precision requires
four times more simulation runs to be performed.

In the case of discrete-time Markov chains (DTMCs), 
the transient stochastic analysis gives the probability distribution over all states of the 
DTMC after $k$ steps. For population models, a step is interpreted as a triggered 
transition. 
The transient solution for DTMCs is the result of $k$ matrix-vector products, and can
be used for the solution of CTMCs, as shown later in Section~\ref{sec:stochastic}.

Numerical analysis tools for discrete-state Markov processes 
such as PRISM\cite{prism04}, INFAMY\cite{infamy}, ETMCC\cite{ETMCC}, MRMC\cite{MRMC}, APNNtoolbox\cite{APNN}, 
SHARPE\cite{sharpe}, SPNP\cite{spnp}, or M\"obius\cite{mobius} have been introduced (see Section~\ref{sec:comparison}). 
However, except for INFAMY, these tools do not accept models with 
possibly infinite state space. 
It is important to note that many population models have an infinite state space, 
that is, the number of reachable states  is infinite. Even when in the real system the number
of molecules, or more generally, individuals is finite, no a priori bound is known, and
models do not include any constraints on the number of molecules, for example in production rules such as
$\emptyset \rightarrow A$. 
Another issue is that existing tools usually implement algorithms that are not 
optimized specifically for population models, and do not scale well on such models.

SABRE is a tool for the transient analysis of Markov population models. 
In other words, SABRE analysis discrete-time, or continuous-time Markov processes  
that have a structured discrete state space and 
 state-depended rate functions. 
In Section~\ref{sec:gcm} we give more details on the space structure 
and the state dependency of rate functions that
are present in Markov processes that represent population models. 

SABRE offers both stochastic and deterministic analysis of population models.
For stochastic analysis, SABRE implements three algorithms: standard uniformization,  
fast adaptive uniformization and Runge-Kutta fourth order method.
The different configurations in which SABRE may operate are depicted in Figure~\ref{fig:software}.
The focus of the tool is on the fast adaptive uniformization method, while the remaining methods are given for completeness and comparison.

Fast adaptive uniformization is a variant of the 
uniformization method\cite{moorsel,stewart} which is,
an efficient method to compute probability distributions if the number of states of the
Markov process is manageable.
However, the size of a Markov process that represents a biochemical 
reaction network is usually far beyond what is feasible.

Fast adaptive uniformization\cite{hibi09} improves the original uniformization method 
at the cost of a small approximation error. 
The main ideas for this improvement are the on-the-fly construction of the 
state space and the restriction imposed on the state space to contain only
states with significant probabilities, e.g. states that have a probability
larger than $10^{-15}$.
Even though fast adaptive uniformization can treat larger models than the previous uniformization methods could, 
as expected, models with remarkably high expected 
populations remain unsolvable and should be studied using deterministic 
analysis of simulation tools. 
A second down side of fast adaptive uniformization is that, due to the approximation
error, it can overlook rare events of the model, e.g. events that occur 
with a very small probability.


SABRE is available on-line at 
\htmladdnormallink{http://mtc.epfl.ch/\textasciitilde{}mateescu/sabre}
{http://mtc.epfl.ch/~mateescu/sabre}.
First, the user gives an input model (either in SBML format or in guarded commands format) 
and a time horizon and than the transient analysis of the system starts (see Figure~\ref{fig:update}).
More details on the usage of the tool are given in Section~\ref{sec:tool}.

\section{Guarded Commands}\label{sec:gcm}

Guarded-command models (GCM) is the input formalism of SABRE. 
GCMs are a textual description of processes and are given
in the style of Dijkstra's guarded-command language\cite{DBLP:journals/cacm/Dijkstra75}.
Their syntax has subsequently been used by languages such as Reactive Modules 
\cite{DBLP:journals/fmsd/AlurH99b} and
by the language for specifying PRISM models\cite{KNP04a}.
The basic unit within GCMs is a transition class, which is expressed as 
a {guarded command} that operates on the state variables of the system.  
A transition class encodes for a possibly infinite number of state transitions.
Within population models, the state variables of
the system are non-negative integers representing numbers of molecules
for each species.  A guarded command takes the form
\begin{verbatim}
       guard |- rate -> update
\end{verbatim}
where the \verb|guard| is a Boolean predicate over the variables 
that determines in which states the corresponding transitions are
enabled. The \verb|update| is a rule that describes the change of the
system variables if the transition is performed.
Syntactically, \verb|update| is a list of statements, each assigning
to a variable an expression over variables.
Assume that \verb|x| is a variable. If, for instance, the update rule
is that \verb|x| is incremented by~$1$, we write \verb|x:=x+1|.  
We assume that variables that are not listed in the update rule do not
change if the transition is taken.  Each guarded command also assigns
a \verb|rate| to the corresponding transitions, which is a function on
the state variables.  Within SABRE, \verb|rate| is given in infix notation. 
In the case of population models, the update function is incrementing or 
decrementing each variable by a constant integer.

For a population model with $m$ reactions, the GCM description is a set of
$m$ guarded commands, which we index as 
\verb|guard|$_j \vdash$ \verb|rate|$_j \rightarrow$ \verb|update|$_j$, 
where each of the commands $j$, with $1 \le j \le m$, describe the $j$-th
reaction of the model. 

GCMs are used to express both CTMCs and DTMCs. The difference between the two 
interpretations comes from the semantic given to the rate function of each command. 
In the case of CTMCs, for a given reaction $j$, the rate function \verb|rate|$_j$
assigns to each state $s$, a positive real value that represents the rate of 
the outgoing transition $j$.

In the case of DTMCs, the rate function \verb|rate|$_j$ assigns to each state $s$, 
a positive real value that represents the transition probability from state $s$ 
to its successor on reaction $j$. 
The functions \verb|rate|$_j$ must define probability distribution over the direct
successor state, that is, for each state $s$ we impose that $\Sigma_j$ \verb|rate|$_j(s) =1$.
If the input is not given in this manner, 
SABRE will automatically normalize the rate functions such that
the probability distribution condition to be fulfilled. Note that this is equivalent to 
interpreting the input as a CTMC and than considering its embedded DTMC.

GCMs are used to model systems that exhibit a finite number of transition types, 
but possibly an unbounded number of states. For example, in a computer 
network, the number of type of events is finite (send message, receive 
message, add node, etc.) but the number of states is countably infinite, 
because it depends on the number of nodes in the network and on the number of 
requests each of them has. The same holds for biochemical reaction networks, 
each reaction type generates a transition class, but the number of states is 
countably infinite, as we do not have any a-priori bound on the variables of the system,
due to productions rules of the type $\emptyset \rightarrow A$. 
We therefore conclude that GCMs are a natural formalism for describing population models\cite{Henz09}

\begin{example} \label{ex:toggle_prm}

The bistable toggle switch is a prototype of a genetic
  switch with two competing repressor proteins and four reactions.
We call the species $A$ and $B$ and we let $x=(x_A,x_B) \in\mathbb N_0^2$ be a vector describing a state of the system. 
The reactions are given in Table~\ref{table:simple_toggle}.
\begin{table}[!t]
\caption{Simple toggle switch example}
\label{table:simple_toggle}
\centering
\begin{tabular}{|l|lll|}
\hline
\textbf{Reaction} & \multicolumn{3}{|l|}{\textbf{Guarded command}}\\
\hline
$\emptyset \to A$ & $\textrm{true}$  &$\vdash  c_1/(c_2+x_B^2)$ &$\rightarrow x_A := x_A + 1$  \\
\hline
$A \to \emptyset$ & $A > 0 $ &$\vdash c_3 \cdot x_1$ &$\rightarrow x_A := x_A - 1$ \\
\hline
$\emptyset \to B$ & $\textrm{true} $ &$\vdash c_4/(c_5+x_A^2)$ &$\rightarrow x_B := x_B + 1$  \\
\hline
$B \to \emptyset$ & $B>0 $ &$\vdash c_6 \cdot x_2$ &$\rightarrow x_B := x_B - 1$ \\
\hline
\end{tabular}
\end{table}

\end{example}

\section {Stochastic and Deterministic Analysis}

SABRE performs a transient analysis of the input system, that is, SABRE computes the state of the system at time $t$
given the state of the system at time $0$. 
SABRE may execute either a stochastic analysis or a deterministic analysis of the input system; and in the
first case the state of the system at time $t$ is actually given as a probability distribution over the discrete 
states of the system. The second type of analysis --the deterministic analysis-- is done over a 
continuous state space, and its result is a single state of this continuous space. 
The result of the deterministic analysis, also known as mean field analysis, is an approximation of the expectation of the stochastic
analysis. 
Each of the two analysis (stochastic and deterministic) may be applied on each of the two semantics (CTMC and DTMC), 
and we will now give short interpretations for the results of the four possible combinations.

\subsection{Stochastic Analysis}\label{sec:stochastic}
\paragraph*{CTMC semantics} We note that the behavior of the CTMC is described as a differential equation (known 
in biochemistry as the chemical master equation) and that $p(t)$ is the solution of that differential equation 
at time $t$. 
The transient stochastic analysis at time $t$, given the initial state $y_0$ with probability $1$, computes
the solution of the chemical master equation at time $t$. 
Within SABRE, the solution $p(t)$ may be computed either by uniformization or by Runge-Kutta explicit fourth order method.

We focus on two uniformization methods for CTMCs, standard uniformization ~\cite{stewart} 
and on its generalization called adaptive unifomization~\cite{moorsel}. 
Standard uniformization splits the given CTMC into a discrete-time Markov chain (DTMC) 
and a Poisson process, whereas adaptive uniformization splits the CTMC into a DTMC and a 
birth process. SABRE implements the optimized algorithm called fast adaptive uniformization 
that has previously been proposed\cite{hibi09}.
One main strength of this algorithm is that it closely tracks the set of significant states of 
the state space, where by a significant state we mean a state with significant probability. 
Secondly, another strength of the fast adaptive uniformization lies in the on-the-fly construction 
of a non-explicit matrix used in the computation of the solution of the DTMC (remember that uniformization
splits the given CTMC into a DTMC and a birth process).

\paragraph*{DTMC semantics} 
DTMC semantics are to be used when the number of triggered transitions, rather than the elapsed time, 
is of interest.  Such situations may arrive, for example, in population genetics models or as a part of 
the uniformization method. 
A transient analysis of a DTMC consists in a series of matrix-vector products:
$$p(k+1) = p(k)\cdot P, $$ with $P$ being the probability transition matrix of the DTMC and 
$p(k)$ being a row vector representing the probability distribution after $k$ steps. 
In our algorithm\cite{hibi09}, the main phase of the DTMC transient analysis is called the propagation phase.
The propagation phase completes the equivalent of a matrix-vector product by moving probability mass from on state $x$ to
all direct successors of $x$ (including $x$ itself if any self-loops are present). 
SABRE approximates the probability distribution 
over the states of the system after $k$ reactions have happened, given
the state $y_0$ of the system before any reaction happens. 
Formally, SABRE approximates the vector $p(k) =\delta_{y_0}\cdot P^k$, where $\delta_{y_0}$ is a dirac probability
distribution in point $y_0$.

\subsection{Deterministic Analysis}
\paragraph*{CTMC semantics} 
We can give an approximate solution of the mean field of the CTMC by 
using the forth order Runge-Kutta method to solve a set of ordinary differential equations simpler than the CME.
This set of equations are known as the reaction rate equations\cite{DBLP:conf/sfm/Gillespie08} and express the change in the 
expectation of each variable over time. 
In the thermodynamic limit (that is, the number of molecules and the volume of the system 
approach infinity) the Markov model and the macroscopic ODE description are equal~\cite{kurtz}.
Therefore, for large populations, the deterministic analysis can be used to approximate the mean field 
of the CTMC.

\paragraph*{DTMC semantics} 
As in the case of CTMCs, for computing the first moment of the transient solution of a DTMC, we can directly 
solve a simpler set of equations that are written directly over variables that represent the expectancies 
of the stochastic solution of the input DTMC model ~\cite{kurtz}.

The expected number of molecules changes deterministically over discrete time, as described by the following equation:
$$x(k+1) = x(k) \cdot A,$$ where $A$ is a probability matrix, and each of its entries $a_{i,j}$ give the probability for
a species $i$ to modify into a species $j$.
Such analysis are useful for discrete-time models as those used to validate communication protocols.

\section{Tool Interface}\label{sec:tool}
\begin{figure*}[!t]
\centering
\includegraphics[width=\textwidth]{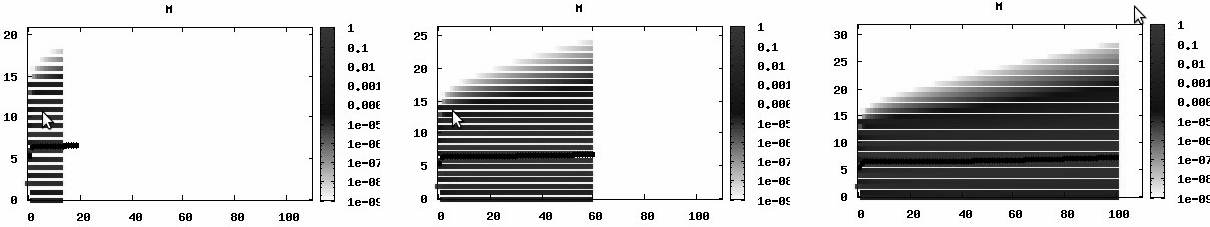}
\caption{Dynamic update of the plots within the web interface. The evoluation of the
probability distribution over the number of monomers in Goutsias' model, as given in\cite{hibi09}.}
\label{fig:update}
\end{figure*}

From the tool's interface, we have several ways of selecting a model for analysis. 
One can load an existing model, upload an SBML file or introduce a GCM text description of the system to analyze. 
SBML is a standardized format for representing models of biological processes, 
such as metabolism or cell signaling and is the input to SABRE's core program. 
GCMs that have update functions with constant increment (or decrement) have a straight forward translation to SBML.

\begin{example} \label{ex:toggle_sbml}
We continue the toggle switch example with its SBML description. For brevity, we only
give one reaction of the model. We observe that the rate function is not restricted to
a particular template and is written following the mathML standard.

\begin{verbatim}
 0 <sbml ...>
 1 <model>
 1 ...
 3 <listOfSpecies>
 4   <species id="A" initialAmount="133"/>
 5   <species id="B" initialAmount="133"/>
 6 </listOfSpecies>
 7 <listOfReactions>
 8   <reaction id="R1">
 9     <listOfProducts>
10       <speciesReference species="A"/>
11     </listOfProducts>
12     <listOfModifiers>
13       <speciesReference species="B"/>
14     </listOfModifiers>
15     <kineticLaw>
16        <math ...>
17          <apply> <divide/>
18            <ci> c1 </ci>
19            <apply> <plus/>
20              <ci> c2 </ci>
21              <apply> <times/>
22                <ci> B </ci>
23                <ci> B </ci>
24              </apply>                                
25            </apply>                                
26          </apply>
27        </math>
28        <listOfParameters>
29        <parameter id="c1" value="3000"/>
30        <parameter id="c2" value="11000"/>
31        </listOfParameters>
32     </kineticLaw>
33   </reaction>
34   ...
35 </listOfReactions>
36 </model>
36 </sbml>
 \end{verbatim}
\end{example}

Once the model is chosen, we choose a configuration of the analysis by choosing the semantics,
the mode and, if needed, the type of stochastic solution. 
Finally, we choose a time horizon, or the number of steps for which we want the system to run. 
We also give as an input a dump time $t_d$, which corresponds to the intermediate results, that is, the system
will compute the distributions for $t_d, 2\cdot t_d, \cdots t$.
The program computes the intermediates and the final results which are then dynamically plotted for each species,
as the computation runs (see Figure~\ref{fig:update}).
If the uniformization method is selected, the user also needs to provide an estimate of the maximal exit rate over
all reachable states. If the estimate is too small, the compuation needs to be restarted, and if the estimate is too 
large, the computation is likely to take longer. It is standard uniformization which is especially touched  
by choosing a too large upper bound on the maximal exit rate. 
Estimating this upper bound by heuristics such as those used for the sliding window algorithm\cite{sliding}
is an on going work.

\section{Software Architecture}\label{sec:archi}

SABRE is available on line, assuring this was a fast and portable release of our implementation. 
The core of our tool is implemented in C++, while the website that hosts it is implemented 
using PHP and Javascript. 
The user provides the desired input through the web interface, than a query is generated to the
3GHz Linux machine on which SABRE is installed. The server sends back to the user intermidiate results
which are then plotted as we show in Section~\ref{sec:tool}.

\subsection {Components}
\begin{figure}[!t]
\centering
\includegraphics[width=3.5in]{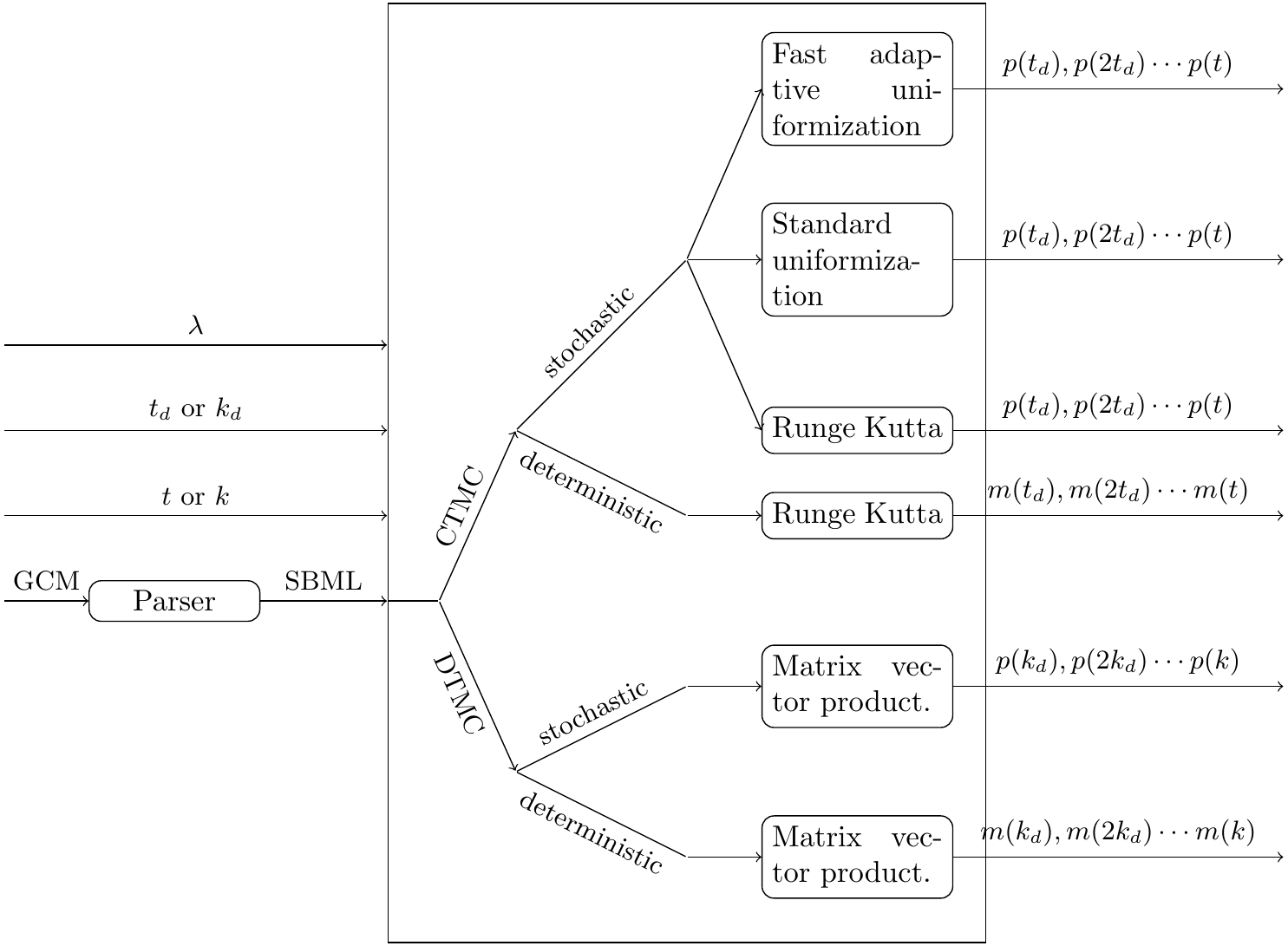}
\caption{Software architecture. Depending on the selected semantics, analysis mode and, eventually, 
type of stochastic solution, SABRE computes the desired results. The vector $p(t)$ is the transient
probability distribution after time $t$, while the vector $p(k)$ is the transient probability
distribution after $k$ steps. For the deterministic analysis, the values $m(t)$ and $m(k)$ correspond
to the mean field of the corresponding CTMC, respectively DTMC. 
The value $\lambda$ represents the maximum exit rate of the CTMC and is required only by uniformization.}
\label{fig:software}
\end{figure}

SABRE's different components are activated as shown in Figure~\ref{fig:software}.
Depending on the chosen semantics, analysis mode and, if necessary, stochastic solution type, 
SABRE calls the coresponding method. 
Some of the functionalities are shared among different methods, for example the DTMC solution
is accessed either directly from choosing the DTMC semantics, either indirectly, by the uniformization
algorithm. 
As well, Runge-Kutta method, is used both as a solver of the CME or as the solver of the
reaction rate equations.

\subsection {Data Structure}

We present an efficient data structure used by SABRE when used in stochastic analysis mode. 
SABRE's main focus is on a fast implementation of the fast adaptive algorithm, so we will use this algorithm to motivate
the choice of our data structure. 
However, the same kind of reasoning works if one wants to optimize the Runge-Kutta implementation. 
The most computationally demanding part of fast adaptive uniformization is the probability propagation phase, 
which performs the equivalent of one matrix-vector product in a DTMC. We therefore need a data structure 
that is efficient during this step.

First, we mention that, for each state, along with the state description, we need to record additional information about the probability of the state,
about its successors, and about the rates/probabilities of the reactions that lead to those respective successors. 
We gather all this information in a structure called \verb|node|. 
During the propagate phase we iterate over all nodes of the state space, and for each node we move probability mass along all 
of its outgoing transitions. Note that, initially the state space has a single state, and that states are dynamically added to
the state space as they are discovered. 
That is, some of the direct successors of $n$ may be newly discovered, and in this case they are added to the state space data structure. 
Therefore, ideally, the data structure used for storing information about the state space would have the following characteristics. 
\begin {itemize}
\item \textbf[Fast sequential access.]  For enumerating all nodes. We note that this is a property of the array primitive type of most programming languages. 
\item \textbf[Fast search.] For quickly finding the successors of a node. We note that this is a property of map or hash type of many programming languages. 
\item \textbf[Fast add.] For dynamically adding newly discovered states to the current state space.
\item \textbf[Fast delete.] For dynamically removing states that have close to zero probability.
\end {itemize}

\begin{table*}[!t]
\caption{Data structure comparison}
\label{table_data_structures}
\centering
\begin{tabular}{|l|l|l|l|l|}
\hline
\textbf{Data Structure} & \textbf{Sequential access} & \textbf{Search} & \textbf{Add} & \textbf{Delete}\\
\hline
Arrays & fast & slow & fast & slow \\
\hline
Maps & slow & fast & fast & fast\\
\hline
Hybrid solution & fast & fast & complex, but fast & complex, but fast\\
\hline
\end{tabular}
\end{table*}

We summarize the comparison between arrays and hashes in Table ~\ref{table_data_structures}. 
Arrays allow fast sequential access, fast add but slow search and delete operations.
Hashes allow fast add, delete, search, but slow sequential access.
We propose a hybrid solution that has the advantages of each data structure at the expense 
of extra memory usage. 

Our hybrid data structure is composed of:
\begin {itemize}
\item array \verb|nodes| that acts as a function from index $\rightarrow$ node
\item hash \verb|index| that acts as a function from state $\rightarrow$ index
\item vector \verb|inactive_nodes| of indices of nodes that have become inactive as a result of a delete. 
\end {itemize}

This mixture of structures lets us give fast implementations for each of the required operations:
\begin {itemize}
\item \textbf{Sequential access} Simple iteration over the elements of \verb|nodes|. 
\item \textbf{Search} Search within \verb|index| followed by an access in \verb|nodes|.
\item \textbf{Delete state} The \verb|nodes| array is allocated statically, so physically 
erasing a node would be expensive. 
The alternative is to mark the node for deletion by inactivating it --setting its 
probability to zero-- and adding it to the \verb|inactive_nodes| vector. 
Because of their zero probability, inactive nodes are ignored when iterating over all states. 
An inactive node has two possible futures: either it will be reoccupied by 
a newly added state, either it will be deleted during a compress phase. 
The compress phase is initiated when the number of inactive nodes covers more then
$20\%$ of the number of both active nodes and inactive nodes and it consists of eliminating all
inactive nodes and rearranging the active nodes in a contiguous region. 
\item \textbf{Add state}  
When we add a state to the state space, we need to assign it to a node within the \verb|nodes| array. 
The \verb|nodes| array is allocated statically and during the program's initialization phase, it is initialized to $2^{20}$ free nodes. 
When we add a new state, if \verb|inactive_nodes| is non-empty, that is, if an inactive node exist, assign the state to this node, 
which now becomes active.
If \verb|inactive_nodes| is empty, we check whether we still have free allocated nodes, that is, we check whether
the number of active nodes has reached the size allocated to \verb|nodes|. If free nodes exist, we assign the new state to a free 
node, if free nodes do not exist we need to allocate extra $2^{20}$ nodes to \verb|nodes| and then pick a newly created free node. 
We note that the reallocation operation is expensive but happens only rarely, e.g. when the state space first reaches one million, two
millions, three millions states and so on.  
\end{itemize}

\section{Case Studies}

We present case studies for stochastic and deterministic analysis of CTMCs and for 
the stochastic analysis of DTMCs. For more and larger experiments on stochastic analysis of 
CTMCs we refer the reader to the paper giving the fast adaptive uniformization algorithm\cite{hibi09}.
All our experiments are performed on a 3GHz Intel Linux PC, with 6 GB of RAM.
We give the results of our experiments in Table~\ref{table:studies}.

\begin{table}[!t]
\caption{Case Studies Summary}
\label{table:studies}
\centering
\begin{tabular}{|c|c||c|c|c|}
\hline
Analysis & Model & Time& Error &States\\
\hline
Stochastic & Exclusive switch &$94s$ & $9e-8$&$3047$\\
\hline
Deterministic & Enzymatic reaction &$<1s$ & $-$ & $1$\\
\hline
Stochastic & Moran's model & $49s$ &$0$&$1001$\\
\hline
\end{tabular}
\end{table}

\subsection {Genetic exclusive switch} 
The exclusive genetic switch we analyze involves two species of proteins
that may bound to the same promoter site. We denote the unbounded 
proteins by $N_1$ and $N_2$ and the bounded ones by $r_1$ and $r_2$\cite{PhysRevE.78.041919}.
The guarded commands for this model are given in Table~\ref{table:toggle}. 
The rate functions are evaluated for the state $(x_{N_1}, x_{r_1}, x_{N_2}, x_{r_2})$,
where $x_{N_1}$ is the number of molecules of type $N_1$ and so on.

\begin{table*}[!t]
\caption{Genetic exclusive switch}
\label{table:toggle}
\centering
\begin{tabular}{|p{0.05in}l|lll|l|}
\hline
\multicolumn{2}{|l|}{\textbf{Reaction}} & \multicolumn{3}{|l|}{\textbf{Guarded Command}} & \textbf{Description}\\
\hline
$\emptyset$&$\to N_1$ & true          & $\vdash g_1 \cdot (1 - x_{r_2}) $           &$ \rightarrow x_{N_1} := x_{N_1} + 1$ & Production of $N_1$\\
\hline
$N_1$&$\to \emptyset$ & $x_{N_1} > 0$ & $\vdash d_1 \cdot x_{N_1}$                  &$ \rightarrow x_{N_1} := x_{N_1} - 1$ &Degradation of $N_1$\\
\hline
$N_1$&$\to r_1$       & $x_{N_1} > 0$ & $\vdash b_1 \cdot (1 - x_{r_1} - x_{r_2} )$ &$\rightarrow x_{N_1} := x_{N_1} - 1; x_{r_1} := x_{r_1} + 1$&Binding of $N_1$\\
\hline
$r_1$&$\to N_1$       & $x_{r_1} > 0$ & $\vdash u_1 \cdot x_{r_1}$                  &$\rightarrow x_{N_1} := x_{N_1} + 1; x_{r_1} := x_{r_1} - 1$& Unbinding of $N_1$\\
\hline
$\emptyset$&$\to N_2$ & true          & $\vdash g_2 \cdot (1 - x_{r_1}) $           &$ \rightarrow x_{N_2} := x_{N_2} + 1$ & Production of $N_2$\\
\hline
$N_2$&$\to \emptyset$ & $x_{N_2} > 0$ & $\vdash d_2 \cdot x_{N_2}$                  &$ \rightarrow x_{N_2} := x_{N_2} - 1$ &Degradation of $N_2$\\
\hline
$N_2$&$\to r_2$       & $x_{N_2} > 0$ & $\vdash b_2 \cdot (1 - x_{r_1} - x_{r_2} )$ &$\rightarrow x_{N_2} := x_{N_2} - 1; x_{r_2} := x_{r_2} + 1$&Binding of $N_2$\\
\hline
$r_2$&$\to N_2$       & $x_{r_2} > 0$ & $\vdash u_2 \cdot x_{r_2}$                  &$\rightarrow x_{N_2} := x_{N_2} + 1; x_{r_2} := x_{r_2} - 1$& Unbinding of $N_2$\\
\hline
\end{tabular}
\end{table*}

When it is bounded to the promotor site, a protein represses the production of the other protein. 
And so, for example, production of $N_1$ only happens if no $N_2$ molecule is bounded to the promoter site (see rate function
of first reaction).
$N_1$ or $N_2$ may bound only to a free promotor site (see rate functions of the third and seventh reaction). 
Note that it always holds that $x_{r_1} + x_{r_2} \le 1$.

\begin{figure}[!t]
\centering
\includegraphics[width=3.5in]{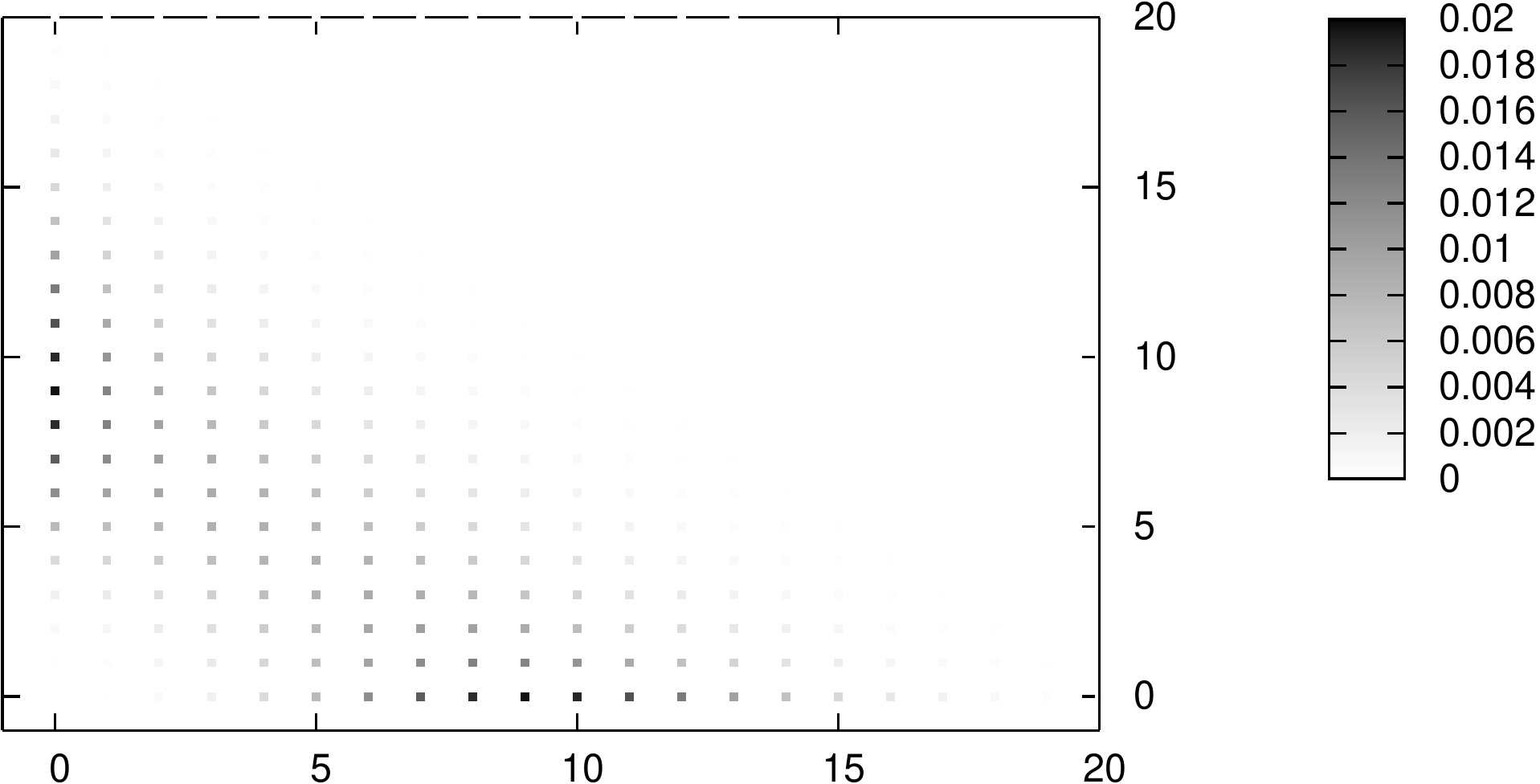}
\caption{Exclusive switch at time 10000. The x-axis gives the number of $N_1$ molecules and the y-axis gives the number of $N_2$ molecules. 
Each point of the plot corresponds to the states of systems that have the have the corresponding number of $N_1$ and $N_2$ molecules. 
The darker the point is, the more probability mass it holds. We can notice the bistable behaviour from the two regions of 
black points, one for $N_1 = 0$ and one for $N_2 = 0$.}
\label{fig:switch}
\end{figure}

We run the system from initial state $(25, 0, 0, 0)$ for a 
period of time of $10000$ units with constants: $g_1 = g_2 = 0.05, d_1=d_2=0.005, b_1=b_2=0.1, u_1=u_2=0.005$,
and present the solution in Figure~\ref{fig:switch}

\subsection {Enzymatic reaction}

We use \emph{enzyme-catalyzed substrate conversion} to exemplify how to
perform a deterministic analysis under continuous-time semantics. 
The enzymatic reaction is described by three reactions (see Table~\ref{table:enzyme}), 
that involve four chemical species, namely, enzyme ($E$),
substrate ($S$), complex ($C$), and product ($P$) molecules.
The state of the system is described by the vector $(x_E, x_S, x_C, x_P)$,
which gives the existing number of molecules of each type. 

\begin{table*}[!t]
\caption{Enzymatic reaction}
\label{table:enzyme}
\centering
\begin{tabular}{|l|lll|l|}
\hline
\textbf{Reaction} & \textbf{Guarded Command} &&& \textbf{Description}\\
\hline
$E+S\to C$ & $x_E>0 \textrm{ and } x_S>0$ &$\vdash c_1\cdot x_E\cdot x_S $ &$ \rightarrow x_E := x_E - 1; x_S := x_S -1; x_C := x_C + 1$ & Formation of complex\\
\hline
$C \to E+S$ & $x_C>0 $ &$\vdash c_2\cdot x_C $ &$ \rightarrow x_E := x_E + 1; x_S := x_S +1; x_C := x_C - 1$ & Dissociation of complex\\
\hline
$C \to E+P$ & $x_C>0 $ &$\vdash c_3\cdot x_C $ &$ \rightarrow x_E := x_E +1; x_C := x_C - 1; x_P := x_P + 1$ & Product production \\
\hline
\end{tabular}
\end{table*}

For our experimental results, we chose the same parameters
as in~\cite{krylovSidje}, that is, initial state $y=(1000,100,0,0)$,
time horizon  $t=70$, and   rate constants $c_{1}=c_{2}=1$ and $c_{3}=0.1$.
For the case deterministic analysis we can not give any error bounds, as shown in Table~\ref{table:studies}.

\subsection {Moran's population model} 
As a simple example of how SABRE operates on DTMC models we choose 
Moran's genetic population model, which can be seen as a set of biochemical
reactions, more specifically as one reversible reaction.

For a population of $N$ individuals, with two alleles, $A_1$ and $A_2$, we are
interested to find the probability of fixation of $A_1$, that is, the probability
for $A_1$ individuals to be equal to $N$ after a certain time. 
We have two reactions: $A_2 \rightarrow A_1$ and $A_1 \rightarrow A_2$. For $x_{A_1}$ individuals with 
$A_1$ allele and $x_{A_2}$ individuals with $A_2$ allele, the probability of the
first reaction is $\frac{1-s}{2} + s\cdot\frac{x_{A_1}}{N}$, 
where $s$ is a small constant. 
As for the second reaction, its probability is $\frac{1-s}{2} + s\cdot\frac{x_{A_2}}{N}$.

We choose $N=1000$ and $s = 2e-3$, the initial state of $x_{A_1} = 1$ and we 
perform a transient analysis until time $k = 10^6$, at this time, 
the probability of fixation is $0.00049$.
In this case the error we obtain is $0$ because no cutting is performed, 
the state space is kept at its complete size of $1001$.

\section{Comparison with other tools}\label{sec:comparison}

Several tools for stochastic analysis of Markov chains have been developed by communities 
such as probabilistic verification, computational biology and performance evaluation among others. 
Here, we provide a comparison with the tools that 
are the closest to SABRE.
The PRISM tool~\cite{prism04}, which is widely used in probabilistic verification,
considers a more general class of Markov processes
than population models. For instance, it does not
restrict the update function such that it allows only a constant 
change of the state variables. The models addressed
by PRISM are less structured and typically they do not have state dependent rate  functions. 
PRISM uses   powerful minimization techniques such as bisimulation
 that do not result in significant reductions in the  
case of population model. PRISM requires that upper bounds on the state 
variables  are given as an input by the user.
As opposed to that the SABRE tool finds appropriate bounds 
automatically and avoids an exhaustive state space exploration.
The drawback is that the SABRE tool cannot guarantee the validity 
of properties such as ``Is the probability to reach state $x$ within 
$t$ time   greater than $p$?''   but gives an approximate solution.
As opposed to that PRISM can guarantee such properties.
On the other hand, since SABRE avoids an exhaustive state space 
exploration it is able to handle much larger models with 
state-dependent rates. 
Infamy is a model-checking tool for infinite-state
CTMCs by Zhang et al.~\cite{infamy}. 
Depending on the desired precision, their algorithm simply explores the 
reachable states up to a finite path depth.
In contrast, our approach takes into account the direction into which the 
probability mass moves, and constructs a sequence of abstract models 
``on-the-fly,'' during the verification process. 
Similar approaches have also been used in the context of biochemical reaction 
networks~\cite{krylovSidje}.

Other tools for stochastic analysis of Markov chains, such as 
ETMCC\cite{ETMCC}, MRMC\cite{MRMC}, APNNtoolbox\cite{APNN},
SHARPE\cite{sharpe}, SPNP\cite{spnp}, and M\"obius\cite{mobius},
are conceived for answering performance analysis questions and 
as PRISM, due to their exhaustive state space exploration
can not be applied to infinite models. 

Dizzy\cite{dizzy}, Snoopy\cite{Heiner08snoopy} and Copasi\cite{copasi} are tools for stochastic simulation alone
and not do not compute probability distributions over states.\cite{CMSB09}
Bio-PEPA\cite{biopepa09} is a language for modeling and analysis of biochemical networks. For numerical analysis and
verification problems Bio-PEPA uses PRISM's engine.

\section{Conclusion}
We have introduced SABRE, a tool for stochastic analysis of biochemical reaction networks and of population models in general. 
We have motivated the choice of guarded commands as input formalism for our tool and the need for a stochastic analyzer
specialized on biological systems. 
SABRE currently has the form of an accessible web tool, which was chosen out of the need to deliver our algorithms 
and optimizations in a fast and portable way. However, an offline version release is planned for the future. 
For completeness and comparison, SABRE also performs deterministic analysis of the input system.

\section*{Acknowledgment}
We thank Marius Mateescu for valuable advices on the web interface and Nick Barton for an introduction to population genetics. 

\bibliographystyle{abbrv}
\bibliography{bio}

\end{document}